\documentclass{PoS}
\usepackage{graphicx}

\title{NLO and NNLO chiral fits for 2+1 flavor DWF ensembles}

\ShortTitle{NLO and NNLO chiral fits for 2+1 flavor DWF ensembles}

\author{\speaker{Robert Mawhinney} \\
       Department of Physics, \\
       Columbia University, \\
       New York, NY \ \ 10027\\
       E-mail: \email{rdm@physics.columbia.edu}}

\author{For the RBC and UKQCD Collaborations}

\abstract{We study the use of NLO and NNLO formulae from SU(2)
chiral perturbation theory to fit results from the 2+1 flavor DWF
QCD ensembles that have been generated by the RBC and UKQCD
collaborations.  These ensembles are at two different lattice
spacings, contain multiple dynamical light quark masses, and include
a variety of partially quenched valence quark masses.  Both NLO and
complete NNLO fits well represent our data, which has $m_\pi$ in
the range 220 to 420 MeV.  With our data, the NNLO fits have NLO
and NNLO contributions of similar size, making the series not
convergent and the extrapolation to physical light quark masses
imprecise.  Thus, we use NLO fit results for our predictions of
$f_\pi$, $f_K$ and the light quark masses.}

\FullConference{The XXVII International Symposium on Lattice Field Theory\\
		 July 26-31, 2009\\
		 Peking University, Beijing, China}

\begin{document}

\section{Introduction}

The RBC and UKQCD Collaborations have been generating 2+1 flavor
domain wall fermion (DWF) QCD ensembles over the last few years.
Extensive results have been published from the first ensemble, which
had two dynamical quark masses, a variety of valence quark masses,
$1/a = 1.72(2)$ GeV and a spatial volume of $(2.75 \; {\rm fm})^3$
\cite{24cubed,24cubed-bk,24cubed-kl3}.  The largest sources of
systematic errors in these results are the $O(a^2)$ errors and
errors from extrapolating from our simulation light quark masses
to the physical light quark masses.  We estimated both of these to
be about 4\% for $f_\pi$, for example.  We now have a second ensemble
at a smaller lattice spacing, which allows us to extrapolate to $a
= 0 $, assuming that our data is in the region where the errors are
of $O(a^2)$.  The second ensemble also has lighter dynamical quarks
and allows us to probe the reliability of the chiral extrapolation
for our data.  This report details our analysis of the combined
data from both ensembles, including chiral and continuum extrapolations.

The details of our ensembles are given in Table \ref{ensembles}.  For the
results from our $1/a = 1.72$ GeV ensembles given in \cite{24cubed},
measurements were made on ensembles of length 3600 MD time units
(after thermalization).  We have generated more lattices, so that
for the $1/a = 1.72$ GeV ensemble with light quark mass $m_l =
0.005$, our results come from 8080 MD time units and for the heavier
light quark mass, $m_l = 0.01$, 7180 time units.  This more than
doubles the measurements from our earlier work.  For the $1/a =
2.32 $ GeV ensemble, we have 6100 MD time units, after thermalization,
for $m_l = 0.004$, 6220 for $m_l = 0.006$, and 5020 for $m_l = 0.008$.

The analysis we present here uses our measurements of light-light
pseudoscalar masses (pions), strange-light pseudoscalar masses
(kaons) and the mass of the $\Omega$ baryon, $m_\Omega$ to set the
lattice scale and determine the physical light and strange
quark masses (we assume $m_u = m_d$ throughout).  As inputs, we
take the known values for $m_\pi$, $m_K$ and $m_\Omega$.  We have
also measured the light-light and light-strange pseudoscalar decay
constants, and predictions for these are an output of our analysis
and a check on our systematic errors.  A major focus of our analysis
is how well full NLO and NNLO chiral perturbation theory formulae fit
our data
and whether the apparent convergence of the series and estimates of
the size of neglected terms are consistent with general theoretical
estimates and known values for $f_\pi$ and $f_K$.


\begin{center}
\begin{table}
\begin{center}
\begin{tabular}{ccccc}
Ensemble & $(m_l^{\rm lat}, \widetilde{m}_l)$ & $m_\pi$ (MeV) &
$(m_{\rm val}^{\rm lat}, \widetilde{m}_{\rm val})$
& $m_\pi^{\rm val}$ (MeV)\\ \hline 
\begin{tabular}{c}
$1/a = 1.72(2)$ GeV  \\
$24^3 \times 64 \times 16$ \\
$(2.75 \; {\rm fm})^3$ \\
$m_h^{\rm lat}= 0.04$ \\
$\widetilde{m}_h= 116.5$ MeV \\
$m_{\rm res}^{\rm lat}= 0.00315$ \\
$m_{\rm res} = 8.5$ MeV
\end{tabular}
&
\begin{tabular}{c}
(0.005, 22.5 MeV) \\
(0.01, 35.5 MeV) 
\end{tabular}
&
\begin{tabular}{c}
328 \\
417
\end{tabular}
&
\begin{tabular}{c}
(0.001, 11.2 MeV) \\
(0.005, 22.0 MeV) \\
(0.010, 35.5 MeV) \\
(0.020, 62.5 MeV) \\
(0.030, 89.5 MeV) \\
(0.040, 116.5 MeV)
\end{tabular}
&
\begin{tabular}{c}
239 \\
328 \\
417 \\
\mbox{} \\
\mbox{} \\
\mbox{} 
\end{tabular}
\\ \hline
\begin{tabular}{c}
$1/a = 2.32(3)$ GeV  \\
$32^3 \times 64 \times 16$ \\
$(2.72 \; {\rm fm})^3$ \\
$m_h^{\rm lat}= 0.03$ \\
$\widetilde{m}_h= 113.1$ MeV \\
$m_{\rm res}^{\rm lat}= 0.000664$ \\
$m_{\rm res} = 2.45$ MeV
\end{tabular}
&
\begin{tabular}{c}
(0.004, 17.2  MeV ) \\
(0.006, 24.6  MeV ) \\
(0.008, 32.0  MeV )
\end{tabular}
&
\begin{tabular}{c}
295 \\
350 \\
397
\end{tabular}
&
\begin{tabular}{c}
(0.002, 9.83 MeV) \\
(0.004, 17.2 MeV) \\
(0.006, 24.6 MeV) \\
(0.008, 32.0 MeV) \\
(0.025, 94.7 MeV) \\
(0.030, 113.1 MeV)
\end{tabular}
&
\begin{tabular}{c}
226 \\
295 \\
350 \\
397 \\
\mbox{} \\
\mbox{} 
\end{tabular}
\\ \hline

\end{tabular}
\end{center}
\caption{\label{ensembles}Run parameters for the simulations presented
here.  Quark and pion masses are given in MeV.
Quark masses are reported in $\overline{\rm MS}(2\; {\rm GeV})$, which
uses the lattice spacings we determine from our analysis and a
separate NPR measurement of the quark mass renormalization factors.
The notation is as in \cite{24cubed} -- in particular, quark masses with
a tilde are total quark masses.}
\end{table}
\end{center}


\section{Observables, reweighting and global fits}

For light-light and heavy-light pseudoscalars, we use Coulomb gauge fixed
wall source propagators.  We fit the propagators starting at 10 lattice
spacings from the wall source and find no dependence on this choice of
fit range.  The plateaus are very long for these states - we fit over 44
lattice spacings for both ensembles.  Coulomb gauge fixed box sources are
used for the $\Omega$, with a box size of $16^3$ for the $1/a = 1.72$ GeV
ensemble and $20^3$ for the 2.32 GeV ensemble.  These sources also give
very good signals, and we fit over a range of 7 lattice spacings.  We
find statistical errors on the pseudoscalar masses and decay constants
in the range of 0.2-0.5\%, and errors on $m_\Omega$ of 0.2-0.7\%.

With these measurements, we then fit our data to SU(2) chiral perturbation
theory (ChPT) formulae.  With SU(2), we do not assume that $m_s$ (or
alternatively $m_K$) is small.  We do need $m_l \ll m_s$.  For the
light-strange sector, we use SU(2) for kaons, as we did in \cite{24cubed}.
In our SU(2) fits, we include $O(a^2)$ corrections to leading order LEC's,
and neglect any $a^2$ dependence for NLO, or higher, LEC's.  The residual
mass effects of DWF are taken into account by including the contribution of
$m_{\rm res}$ in the total quark mass.

Of course, there are many ways to write down SU(2) ChPT formula at NLO,
since any rearrangement of the series that only changes it at NNLO is
equally valid.  Since we have no {\em a priori} reason to know which
particular ordering is the most convergent for a particular quantity,
and it may seem unlikely that any one reordering will be optimal for
all quantities, we use the series as an expansion in $f$, the chiral
limit value for the pion decay constant.  The pseudoscalar masses that
enter at NLO and higher are just $m^2 = 2 B m_q$, the leading order
expressions.  This view of the series also readily extends to the full,
continuum NNLO forms, as given in \cite{bijnens}.

Since the total lattice quark mass enters into the ChPT expressions
and we are working at two different lattice spacings, we need a
renormalization factor to relate bare quarks at one lattice spacing
to the other.  We have gotten this ratio three different ways:  1)  from
NPR calculations at the two different lattice spacings, 2)  from matching
the two lattice spacings at unphysical quark masses \cite{kelly} and 3)
from a global fit where the ratio is a free parameter and it is fit for
in an overall $\chi^2$ minimization step.  All three methods agree within
their errors.

A last issue in our global fits is the difference between the
dynamical heavy quark mass in the simulations, $m_h$, and the
physical strange quark mass, $m_s$.  When the simulations are run,
the value of $m_h$ to use is not known.  Only after a complete
analysis of the data does one gain knowledge about the correct
value of $m_h$.  In SU(2) ChPT, unlike SU(3), all LEC's are implicit
functions of $m_h$ and cannot be extrapolated to the physical value.
(For valence heavy quark mass dependence, it is easy to interpolate
between the values for a hadron mass that are measured with different
valence quark masses to a self-consistently determined strange quark
mass.)  By reweighting our observables from the simulated $m_h$ to the
desired $m_s$, we can remove any systematic error from the (generally
mild) dependence on $m_h$.

\begin{center}
\begin{figure}
\begin{minipage}{1.25\textwidth}
\begin{center}
\hspace{-1.6in}
\includegraphics[width=0.40\textwidth,viewport=0 180 575 625,clip]{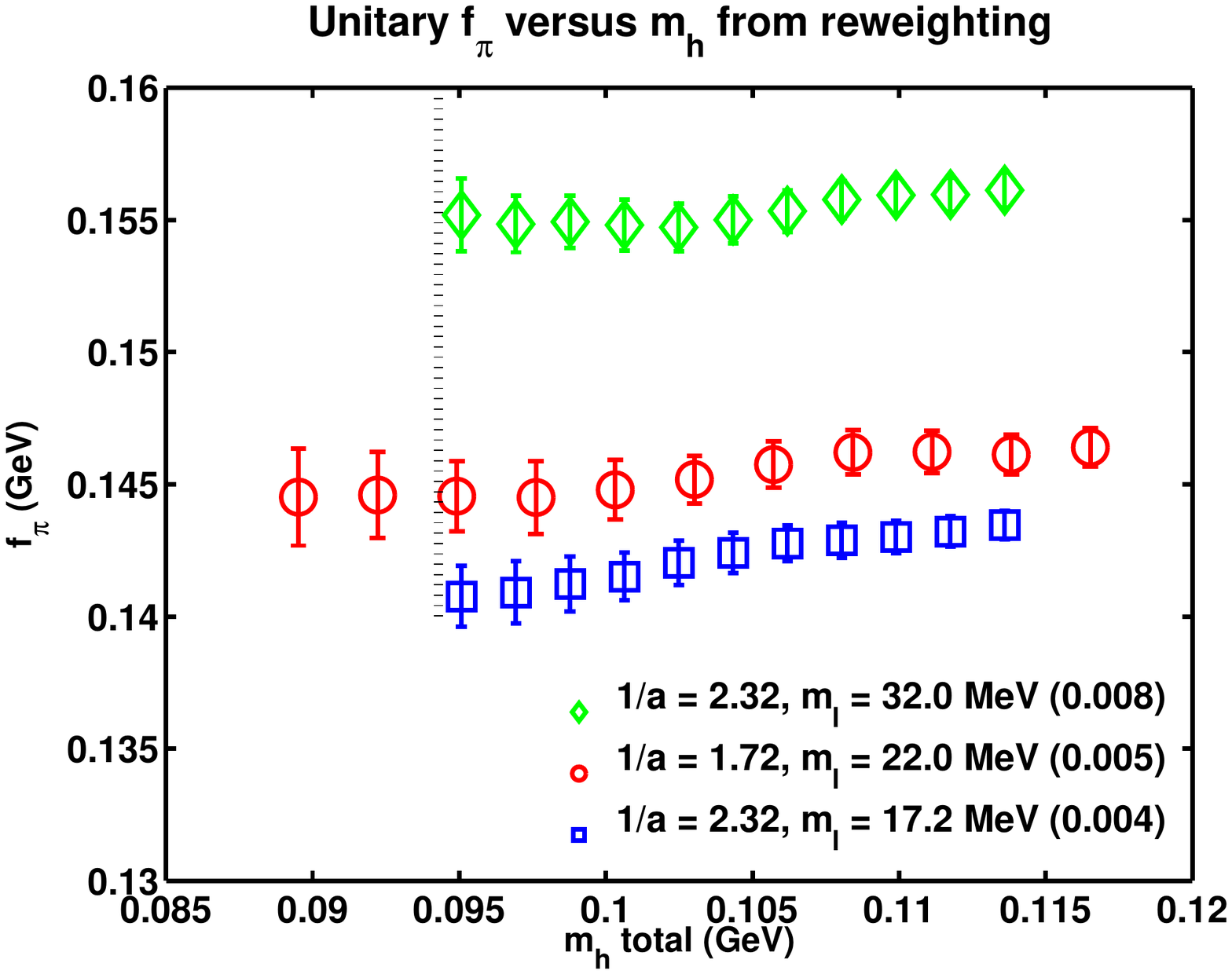}
\includegraphics[width=0.40\textwidth,viewport=0 180 575 625,clip]{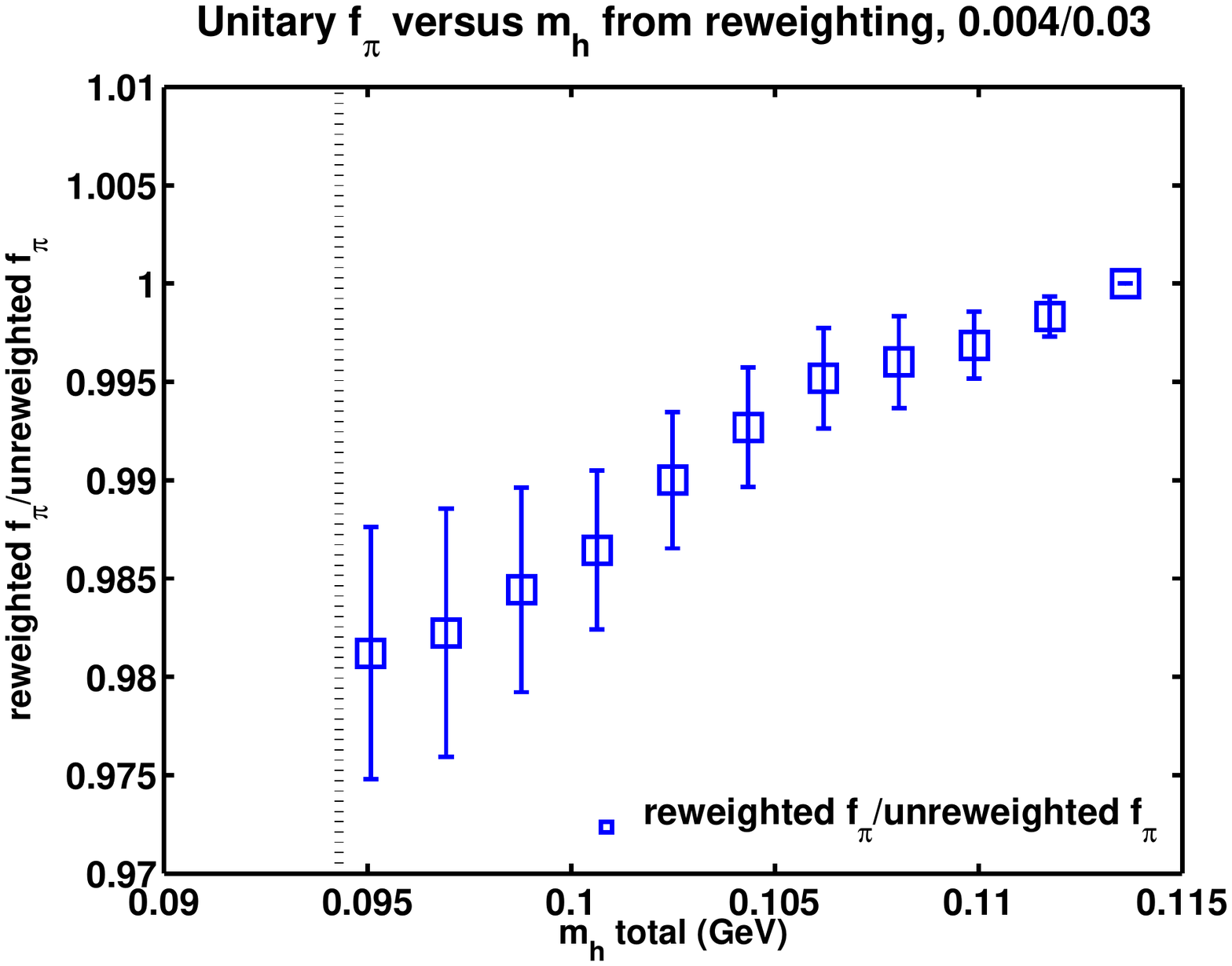}
\end{center}
\end{minipage}
\hspace{-1.4in}
\caption{\label{fig:fpi_rw}
The left graph shows the reweighted value for $f_\pi$ for some of our
ensembles.  The right graph is the ratio of $f_\pi$, reweighted to the
quark mass given on the horizontal axis, to the unreweighted $f_\pi$.}
\end{figure}
\end{center}

The left graph in Figure \ref{fig:fpi_rw} shows the unitary
values for $f_\pi$ as a function of the reweighted dynamical heavy
quark mass for three of our ensembles.  The right graph shows the
ratio of the reweighted value of $f_\pi$ to the unreweighted value
for the $1/a = 2.32 $ GeV, 0.004/0.03 ensemble.  Four stochastic
estimators per mass step were used here and studying the reweighting
dependence versus the number of stochastic hits indicates that four
are sufficient.  We see a clear signal for a small dependence on
the strange quark mass.  More details of this procedure are also given in
\cite{chulwoo}.

Three different global fits, using the above procedure have been performed
by members of our collaboration and the agreement for physical quantities
is very good.  Two types of fits use results from matching the lattices
at unphysical quark masses to constrain the ratio of quark renormalization
factors and/or lattice spacings.  A third fit self-consistently fits the
data, determining the ratio of lattice spacings, quark mass
renormalizations and LECs that give the best $\chi^2$.  We do uncorrelated
fits to our data, since we find our covariance matrices are very singular,
due to the data being strongly correlated.  As shown in \cite{dawson} in
such a case an uncorrelated fit gives the correct answer, but the quoted
$\chi^2$ is not a reliable goodness-of-fit indicator. 

\begin{center}
\begin{figure}
\begin{minipage}{1.25\textwidth}
\begin{center}
\hspace{-1.6in}
\includegraphics[width=0.40\textwidth,viewport=0 180 575 625,clip]{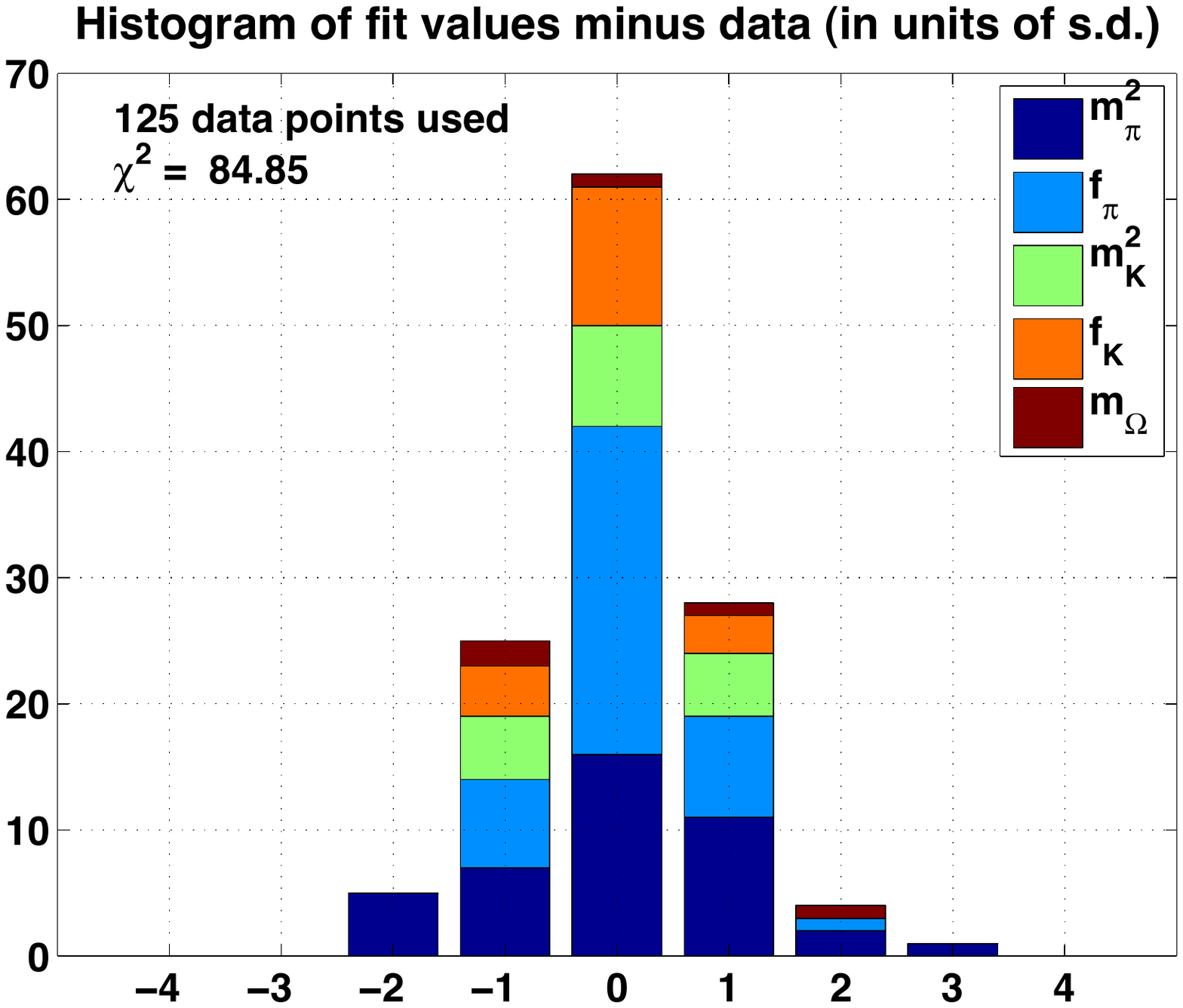}
\includegraphics[width=0.40\textwidth,viewport=0 180 575 625,clip]{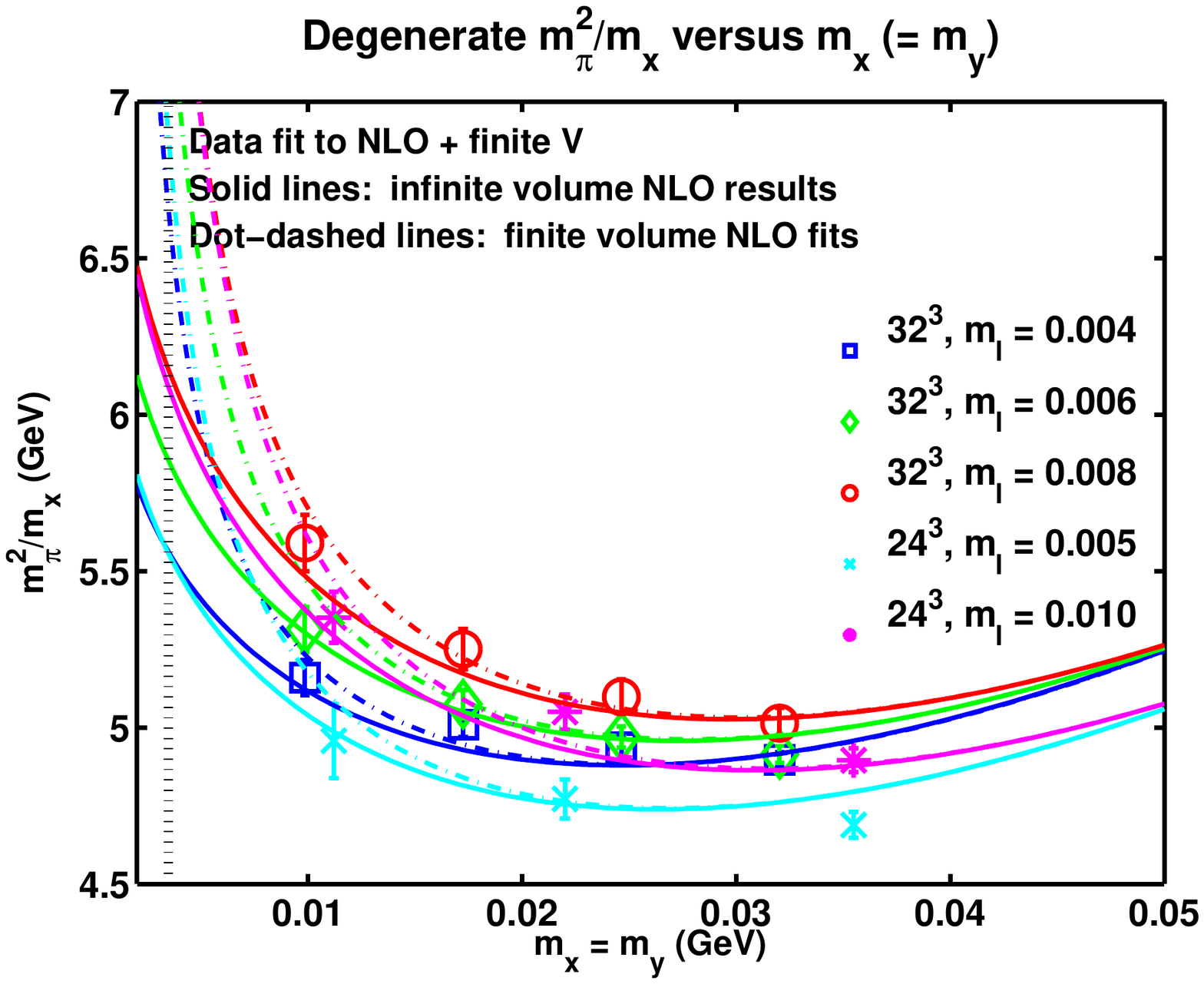}
\end{center}
\end{minipage}
\caption{\label{fig:chi-sq}
The left graph is a histogram of the deviations of fit values from
measured values, in units of the $\sigma$ for the data points.
The right graph shows results from the partially quenched, NLO ChPT
fits for $m_\pi^2$.}
\end{figure}
\end{center}

The left graph of Figure \ref{fig:chi-sq} shows a histogram of the
uncorrelated $\chi^2$ for our fits, which involve 125 data points
and about 20 parameters, depending on precisely which fit was done.
All of our partially quenched data for light-light and heavy-light
pseudoscalars and the $\Omega$ are fit simultaneously, under an
outer jackknife loop, which produces the errors.  The fit is to NLO
order in ChPT, including finite volume effects.  The physical values
for $m_\pi$, $m_K$ and $m_\Omega$ are used to determine the lattice
scale and quark masses.  From NPR \cite{yaoki}, we use a value of
$Z_m = 1.590$ for the $1/a = 2.32$ GeV ensemble to convert
lattice quark masses into continuum masses renormalized in
$\overline{\rm MS}(2\; {\rm GeV})$.  The histogram shows that the
fits are in good agreement with the error bars on each data point.
The right graph of Figure \ref{fig:chi-sq} shows the comparison
between the NLO ChPT fit results and the data for pions made of
degenerate quarks.  Curvature consistent with the expected chiral
logarithms is seen.

\begin{center}
\begin{figure}
\begin{minipage}{1.25\textwidth}
\begin{center}
\hspace{-1.6in}
\includegraphics[width=0.40\textwidth,viewport=0 180 575 625,clip]{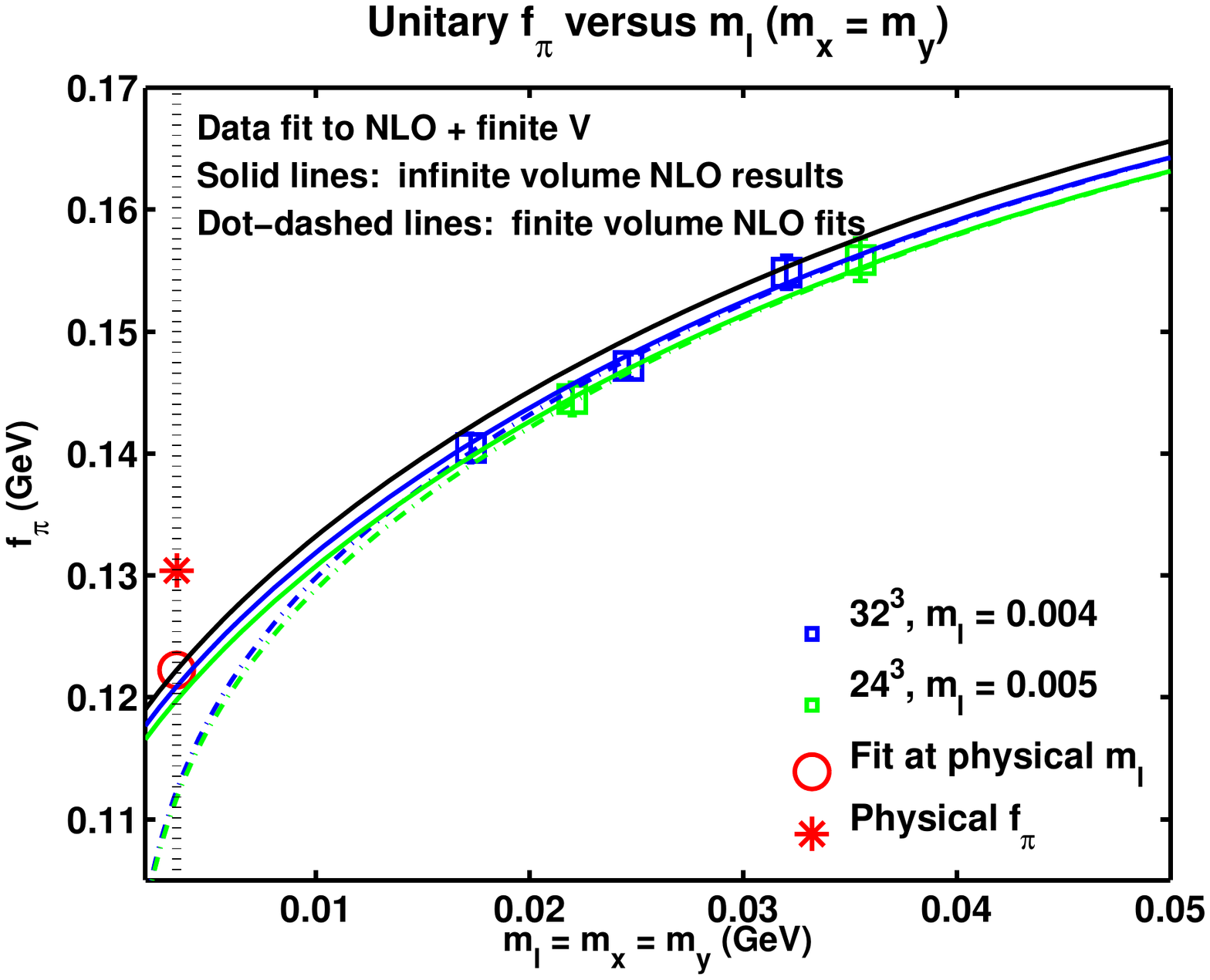}
\includegraphics[width=0.40\textwidth,viewport=0 180 575 625,clip]{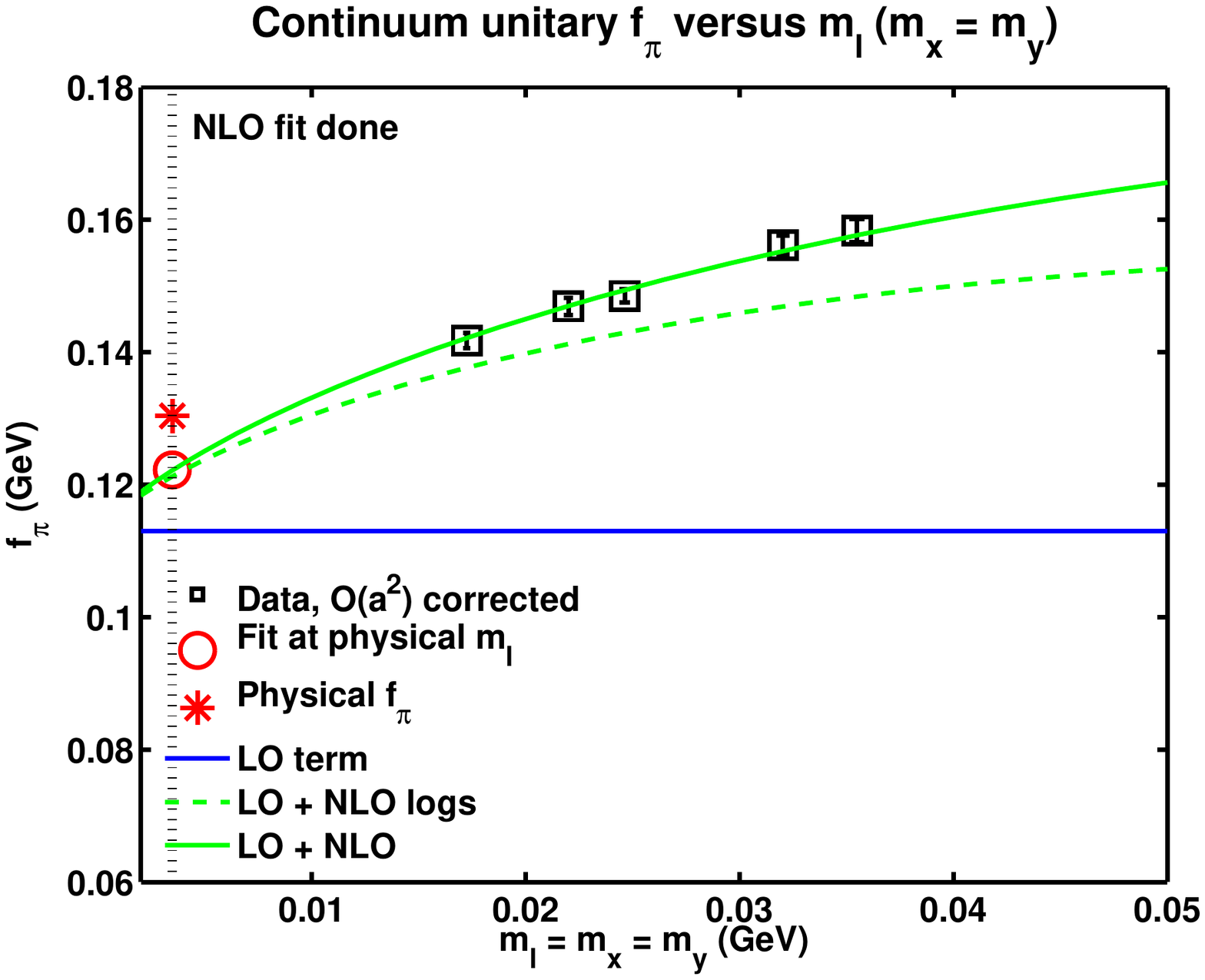}
\end{center}
\end{minipage}
\caption{\label{fig:fpi_unitary}
The left graph shows the unitary, light-light pseudoscalar decay constant
versus quark mass, from our NLO fit.  The right graph shows the contribution
of various terms to the fit.}
\end{figure}
\end{center}

Figure \ref{fig:fpi_unitary} shows our results for the unitary
light-light pseudoscalar decay constant and our fits.  We find
$f_\pi = 122.2 \pm 3.4_{\rm stat} \; {\rm MeV}$.  The right graph
shows the LO and NLO contribution to the fit and one sees that in
the region where we have data, the NLO contribution is a 20-30\%
correction to the LO result.  From this one would expect an NNLO
error of order $0.2^2$ to $0.3^2$ or $O(4-9\%)$.  One sees that our
prediction for $f_\pi$ is low by roughly this amount.  We also
estimate a similar ChPT systematic effect by looking at simple,
analytic fits to our data \cite{kelly}.  We quote a preliminary ChPT
systematic error given by the square of the ratio of
NLO to LO in the lightest quark mass region where we have data.

{\em A priori}, one has little information about the size of NNLO
ChPT contributions.  If the series is reasonably convergent when
the next order is added, the NNLO terms should be roughly the
square of the NLO contributions.  Of course ChPT differs from a
renormalizable field theory in that many new LEC's enter at the
next order, as well as contributions from LEC's at the current order
times logarithms.  For DWF, where we have continuum chiral symmetries
at finite lattice spacing, we can fit our data to the continuum NNLO SU(2)
formulae, to help address these questions.


\section{Full NNLO fit}

We have also fit our data to the full, continuum NNLO ChPT results
for SU(2) given by Bijnens and Lahde \cite{bijnens}.  This adds 13
new parameters to our fits, five $L_i$ and eight, linearly-independent
combinations of the $K_i$.  For now, we have done NNLO fits keeping
the lattice spacing and ratio of quark mass renormalization factors
fixed to the values returned from NLO fits.

\begin{center}
\begin{figure}
\begin{minipage}{1.25\textwidth}
\begin{center}
\hspace{-1.6in}
\includegraphics[width=0.40\textwidth,viewport=0 180 575 625,clip]{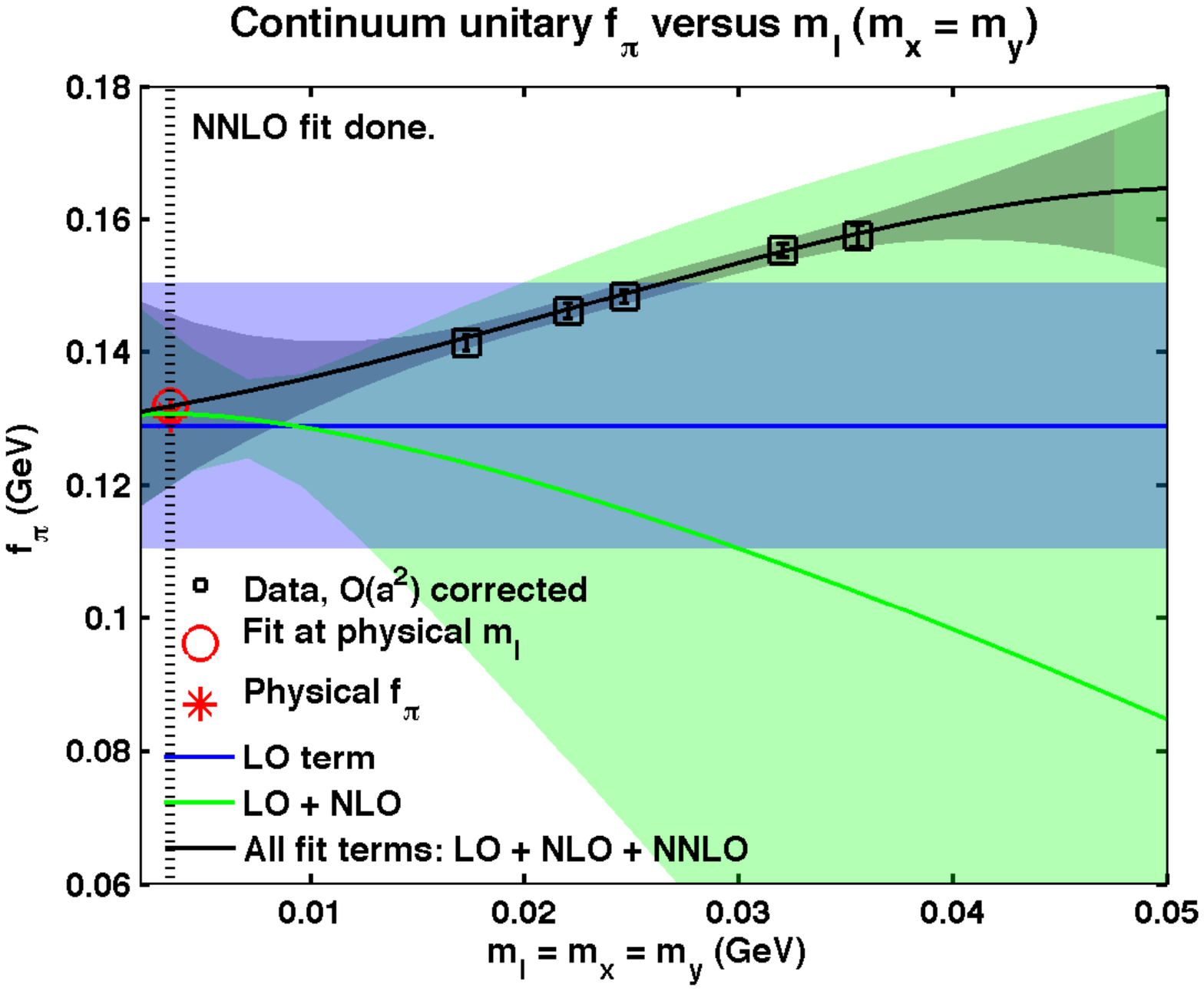}
\includegraphics[width=0.40\textwidth,viewport=0 180 575 625,clip]{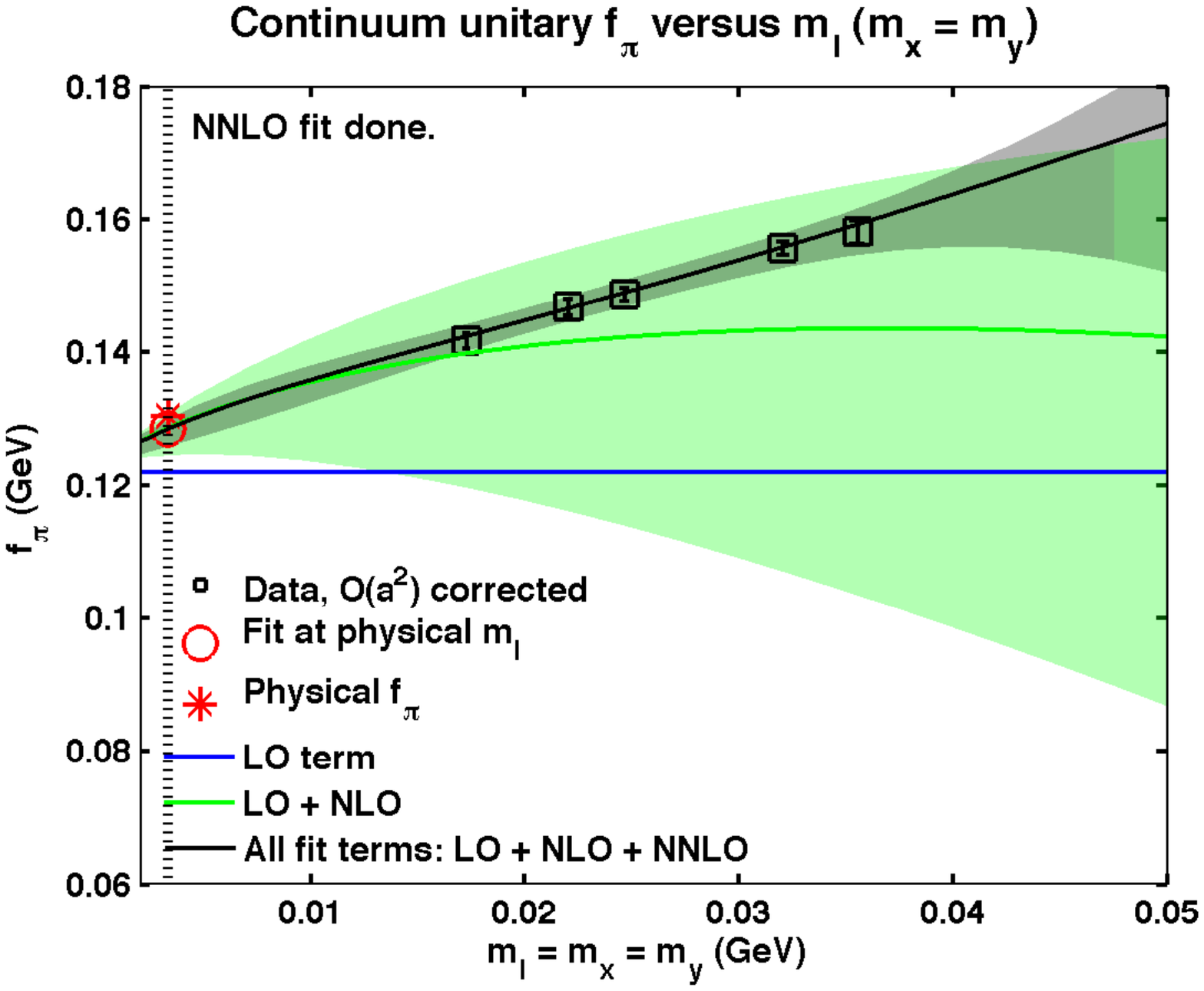}
\end{center}
\end{minipage}
\caption{\label{fig:nnlo}
The left graph is $f_\pi$ from a full NNLO fit to our data. The right graph
is also for a full NNLO fit, but with the constraint that the
SU(2) chiral limit decay constant, $f = 122.0$ MeV.}
\end{figure}
\end{center}

The left graph in Figure \ref{fig:nnlo} shows the results for $f_\pi$
from a full NNLO fit to all of our partially quenched, light-light
masses and decay constants.  Both lattice spacings are fit
simultaneously using a standard, least-squares approach and the
uncorrelated $\chi^2 = 21.8$ with 125 data points and 33 parameters.
Statistical errors come from a jackknife analysis.  The NNLO fit
predicts $f_\pi = 133 \pm 13_{\rm stat}$ MeV, $f = 130.4 \pm 20.0_{\rm
stat}$ and gives values for $m_{ud}$ and $m_s$ the same, within
statistical errors, as the NLO fits.  The blue band in the figure
gives the one $\sigma$ error for the LO contribution, the green
shows the error for the LO + NLO contribution and the black the
error for the LO + NLO + NNLO contribution.  The large size of the
green error band means that, for the majority of our fits under the
jackknife loop, the size of the NNLO contributions is very large
and the series is not convergent.  This observation, along with
the increase in the statistical error for $f_\pi$ from the extra
degrees of freedom in the fit, means that we cannot get an
accurate extrapolation to physical light quark masses by fitting
our data, which has $m_\pi = 220$ to 420 MeV, to NNLO order in ChPT.

The right graph in Figure \ref{fig:nnlo} shows the results for
$f_\pi$ from a full NNLO fit, with the constraint that the SU(2)
chiral limit decay constant, $f = 122$ MeV.  (This value of $f$
comes from the phenomenological value for $\bar{l}_4$ and has an
uncertainty of about 1 MeV.) The total, uncorrelated $\chi^2$ for
these fits is about 25 and gives $f_\pi = 127.9 \pm 1.8_{\rm stat}$
MeV.  (The smaller statistical error for this fit comes from the
strong constraint in the chiral limit.) However, the large error
on the LO + NLO contribution, means that, in a $\sim 25$\% of our fits,
we have very large NNLO contributions.  Thus we cannot demonstrate
a reasonably convergent series at this point, even with a constraint.


\section{Summary and Conclusions}

We have found that our 2+1 flavor DWF QCD data from two lattice
spacings, with partially quenched $m_\pi$ from 220 to 420 MeV,
can be fit with either NLO or full NNLO ChPT formulae.  Standard,
least-squares fits yield small, uncorrelated values for $\chi^2$,
so the fit formulae well represent our data.  For unconstrained
NNLO fits, we find $f_\pi = 133 \pm 13_{\rm stat}$ MeV and the
series is not convergent.  Constraining the chiral limit
value for $f$ to be 122.0 MeV gives $f_\pi = 127.9 \pm 1.8_{\rm stat}$
and the series appears less poorly convergent.  With the
current data set, NNLO fits do not provide a reliable extrapolation
to physical light quark masses.

Turning to NLO fits, we find the NLO terms for $f_\pi$ are 20-30\%
the size of the LO term, at quark masses where we have data.  From
this, we estimate a 4-9\% correction due to NNLO terms and our value
for $f_\pi$ deviates from the physical value by about this much.
We take the square of the fractional size of the NLO correction as
an estimate of our systematic ChPT error.  We also have used NPR
to determine the quark mass renormalization, allowing us to determine
values for $m_{ud}$ and $m_s$.  Our preliminary results are:
\begin{center}
\begin{tabular}{cc}
  $ m_{ud}^{\overline{\rm MS}}(2 \; {\rm GeV})  =  3.47 \pm 0.10_{\rm
  stat} \pm 0.17_{\rm NPR} \; {\rm MeV } $  \\
  $ m_{s}^{\overline{\rm MS}}(2 \; {\rm
  GeV})  =  94.3 \pm 3.4_{\rm stat} \pm 4.5_{\rm NPR} \; {\rm MeV}$ &  
  $m_{s}/m_{ud}  =  27.19 \pm 0.35_{\rm stat}$ \\ \hline
  $f_\pi  =  122.2 \pm 3.4_{\rm stat} \pm 7.3_{\rm ChPT} \; {\rm MeV} $ \\
  $f_K  =  149.7 \pm 3.8_{\rm stat} \pm 2.0_{\rm ChPT} \; {\rm MeV} $ & 
  $f_K/f_\pi  =  1.225 \pm 0.012_{\rm stat} \pm 0.014_{\rm ChPT}$ \\ \hline
  $f  =  113.0 \pm 3.8_{\rm stat} \pm 6.8_{\rm ChPT} \; {\rm MeV}$ \\
  $f_K^{(0)}  =  144.8 \pm 4.2_{\rm stat} \pm 2.0_{\rm ChPT} \; {\rm MeV}$ &
  $f_K^{(0)}/f  =  1.282 \pm 0.015_{\rm stat} \pm 0.017_{\rm ChPT}$ 
\end{tabular}
\end{center}

This work and the author were supported in part by US DOE grant
\#DE-FG02-92ER40699.  We thank Johans Bijnens for his Fortran code
for the evaluation of NNLO SU(2) ChPT.  The RBC and UKQCD collaborations
receive additional support from the US DOE, the RIKEN-BNL Research
Center, and PPARC in the UK.  Part of this work used computer time
granted by the USQCD Collaboration.  We thank RIKEN, BNL, the US
DOE and PPARC in the UK for providing facilities essential to this
work.  Producing the computationally demanding $1/a = 2.32$ GeV ensembles
has only been possible through the transformative scale of resources
made available from the Argonne Leadership Class Facility.


\end{document}